\documentclass[preprint,12pt]{elsarticle}
\journal{Mechanics of Materials}
\usepackage{graphics,epsfig}
\usepackage{graphicx}
\usepackage{float}
\usepackage{amssymb,amstext,amsmath}
\usepackage{mathtools}
\usepackage{booktabs}
\usepackage{natbib}
\usepackage[latin1]{inputenc}
\usepackage{epstopdf}
\usepackage{physics}
\usepackage{multirow}
\usepackage{diagbox}
\begin{document}

\begin{frontmatter}
\title{Thermodynamic dislocation theory: Application to bcc-crystals}
\author{K. C. Le$^{a,b}$\footnote{Corresponding author: lekhanhchau@tdtu.edu.vn.}, S. L. Dang$^c$, H. T. Luu$^c$, N. Gunkelmann$^c$}
\address{$^a$Materials Mechanics Research Group, Ton Duc Thang University, Ho Chi Minh City, Vietnam. 
\\
$^b$Faculty of Civil Engineering, Ton Duc Thang University, Ho Chi Minh City, Vietnam.
\\
$^c$Clausthal University of Technology, Institute of Applied Mechanics, Adolph-Roemer Str. 2A, Clausthal Zellerfeld 38678, Germany.}
\begin{abstract} 
This paper presents the thermodynamic dislocation theory containing several modifications over its first version which was originally proposed by Langer, Bouchbinder, and Lookman (2010). Employing a small set of physics-based material parameters identified by the large scale least squares analysis, we show that the theory can fit the stress-strain curves of bcc crystals niobium, tantalum, tungsten, and vanadium over a wide range of temperatures and strain rates. 
\end{abstract}

\begin{keyword}
modelling/simulations \sep other materials \sep plasticity methods \sep dislocations \sep plasticity.
\end{keyword}
\end{frontmatter}

\section{Introduction}
\label{sec:Introduction}
For more than eight decades, physicists, materials scientists and engineers have tried unsuccessfully to search for a {\it predictable} dislocation-based plasticity. Commonly accepted approaches in crystal plasticity (see \cite{Johnson1983,Zerilli1987,Kocks2003} and the references therein) are phenomenological in nature: one proposes constitutive equations based on the hardening or softening behavior of the stress-strain curves of ductile metals and alloys, sometimes even without using dislocation mechanisms to explain them. As a result, the theories are rather descriptive than predictive, and one should change the constitutive equations as well as the material parameters when, say, the loading condition, the grain size, the strain rate, or the temperature are changed. The question of whether a predictive dislocation-based plasticity is feasible remains a matter of serious debate. The prevailing opinion among experts in the field is that it is not. For example, Cottrell \cite{Nabarro2002} has argued that ``strain hardening (rather than turbulence) is the most difficult remaining problem in classical physics.''

The real progress has recently been achieved in the theory of dislocation mediated plasticity  proposed by Langer, Bouchbinder, and Lookman \cite{LBL-10}, called LBL-theory for short. The breakthrough therein consists in decoupling the system of dislocated crystal into configurational and kinetic-vibrational subsystems. The configurational degrees of freedom describe the relatively slow, i.e. infrequent, atomic rearrangements that are associated with the irreversible movement of dislocations; the kinetic-vibrational degrees of freedom the fast vibrations of atoms in the lattice. The governing equations of LBL-theory involving the effective disorder temperature are derived from the kinetics of thermally activated dislocation depinning and irreversible thermodynamics of driven systems. This LBL-theory has been successfully used to simulate the stress-strain curves for copper over fifteen decades of strain rate, and for temperatures between room temperature and about two third of the melting temperature showing the excellent agreement with the experiments conducted by Follansbee and Kocks \cite{FK-88}. The theory has been extended to include the interaction between two subsystems \cite{Langer2017} and used to simulate the stress-strain curves for aluminum and steel alloy \cite{LTL-17} which exhibit thermal softening in agreement with the experiments conducted in \cite{Shi1997,Abbod2007}. It was again extended and employed to predict the formation of an adiabatic shear band in rapidly loaded HY-100 steel \cite{LTL-18} that shows quantitative agreement with the experimental observations by Marchand and Duffy \cite{Marchand1988}. In \cite{LLT-20,Le-20} one of us has shown that some modifications concerning the theory as well as the method of identification of material parameters are required to capture the behavior of real fcc-polycrystals (see also \cite{LL-20,LB-20}). The LBL-theory, together with these modifications, is called thermodynamic dislocation theory (TDT). However, it still remains unclear whether this theory can be applied to other polycrystals like bcc or hcp. The aim of this paper is to show that TDT is applicable to bcc-polycrystals as well. 

To achieve this aim, we begin in Section~\ref{TDT} with a discussion of LBL-theory and its necessary modifications leading to thermodynamic dislocation theory applicable to polycrystals under tension/compression. Next, in Section~\ref{Rescaled} we present the rescaled governing equations of the TDT, which are more convenient for numerical simulations. In Section~\ref{DATA} we show the data for niobium \cite{Nemat-Nasser2000}, tantalum \cite{Nemat-Nasser1997}, tungsten \cite{Lennon2000} and vanadium \cite{Nemat-Nasser2000a}, describe our methods for using these data to determine the material specific parameters and describe our theoretical interpretation of these measurements.  We conclude in Section \ref{Conclusions} with some remarks on the significance of these calculations.

\section{Thermodynamic dislocation theory}
\label{TDT}

Let us start with the kinetics of thermally activated dislocation depinning leading to the following formula for the plastic slip rate of each slip system (see \cite{LBL-10,Le2020} for the derivation)
\begin{equation}
\label{eq:plastic-slip-rate} 
\dot{\beta}\equiv \frac{q}{t_0}=\frac{b}{t_0}\sqrt{\rho} \exp \Bigl[-\frac{T_P}{T} e^{-\tau/\tau_T(\rho)} \Bigr] .
\end{equation}
Here, $t_0$ is a microscopic time characterizing the depinning rate, $b$ the Burgers vector, $\rho$ the dislocation density, $T$ the ordinary temperature, $T_P$ the pinning energy barrier at zero stress (in the temperature unit), and $\tau_T(\rho)=\mu_T b \sqrt{\rho }$ the Taylor stress. There is still an ongoing debate in the materials science community about the dislocation mechanism that controls the rate of plastic slip in bcc crystals (see, e.g., \cite{Nemat-Nasser2000} and the references therein). In this paper we will assume that the kinetic equation \eqref{eq:plastic-slip-rate} remains valid for the bcc-crystals independent of the depinning mechanism. Averaging this equation over all positively oriented slip systems under the assumption that their orientations are equally probable \cite{LLT-20}, and taking into account that $\bar{\tau}=2\sigma/3\pi$, with $\sigma$ being the macroscopic stress, and $\dot{\bar{\beta}}t_0=\frac{2}{\pi}\dot{\varepsilon}^pt_0$, with $\varepsilon^p$ being the macroscopic plastic strain, for polycrystals subjected to tension/compression, we obtain
\begin{equation}
\label{eq:plastic-strain-rate} 
\dot{\varepsilon}^p\equiv \frac{\pi}{2}\frac{q(\sigma,\rho,T)}{t_0}=\frac{\pi}{2}\frac{b}{t_0}\sqrt{\rho }\exp \Bigl( -\frac{T_P}{T}e^{-\sigma /\bar{\mu}_Tb\sqrt{\rho}} \Bigr) ,
\end{equation}
with $\bar{\mu}_T=3\pi \mu_T/2$. Formula \eqref{eq:plastic-strain-rate} is the first modification of LBL-theory that follows from the averaging procedure. Another not less important modification is that we abandon the rather ad-hoc assumption $t_0=10^{-12}$s and will identify it later (see Section~\ref{DATA}). 

The next modification concerns the equation for the stress. Using the averaged resolved shear stress $\bar{\tau}=2\sigma/3\pi$, the averaged elastic resolved shear strain $\bar{\varepsilon}^e_{sm}=2(1+\nu)(\varepsilon-\varepsilon^p)/3\pi $, with $\nu $ being Poisson's ratio, as well as the average constitutive equation $\bar{\tau}=2\mu \bar{\varepsilon}^e_{sm}$ \cite{LLT-20}, we obtain
\begin{equation}
\label{eq:stress}
\sigma =2\mu (1+\nu)(\varepsilon-\varepsilon^p).
\end{equation}
Taking the time derivative of \eqref{eq:stress}, replacing $\dv{t}=(q_0/t_0)\dv{\varepsilon}$ (which is due to the constant strain rate), and recalling \eqref{eq:plastic-strain-rate}, we arrive at
\begin{equation}
\label{eq:stressrate}
\dv{\sigma}{\varepsilon}=2\mu (1+\nu)\Bigl( 1-\frac{q}{\bar{q}_0}\Bigr),
\end{equation} 
where $\bar{q}_0=2q_0/\pi$. We see that, in contrast to LBL-theory, the ``geometric factor'' for polycrystals under tension/compression must be $\alpha=2(1+\nu)$ (instead of 1).

The equations for the dislocation density $\rho $ and the configurational temperature $\chi $ should be derived from the first and second law of thermodynamics of configurational subsystem. These equations remain the same as those derived in \cite{LBL-10}. They read
\begin{align}
\dv{\rho}{\varepsilon}&=\frac{\kappa_\rho }{\bar{\mu}_T \zeta ^2(\rho,\bar{q}_0,T)b^2} \frac{\sigma q}{\bar{q}_0}\Bigl[1-\frac{\rho}{\rho_s(\chi)}\Bigr] , \label{eq:15}
\\
\dv{\chi}{\varepsilon}&=\frac{\kappa_\chi }{\bar{\mu}_T}\frac{\sigma q}{\bar{q}_0} \Bigl(1-\frac{\chi}{\chi_0}\Bigr) . \notag
\end{align}
Here, $\rho_{s}(\chi)=\frac{1}{a^2}e^{-e_d/\chi }$, with $e_d$ being the formation energy for dislocations and $a$ a length scale in the order of the atomic spacing, corresponds to the most probable (steady-state) dislocation density at fixed configurational temperature $\chi$, while
\begin{displaymath}
\zeta(\rho,\bar{q}_0,T)=\ln \Bigl( \frac{T_P}{T}\Bigr)-\ln\Bigl[\ln\Bigl(\frac{b\sqrt{\rho}}{\bar{q}_0}\Bigr)\Bigr] .
\end{displaymath}
However, there is an important modification concerning the factor $\kappa_\chi $. If the latter would be independent of temperature, the initial configurational temperatures prove to be very close to the steady-state effective temperature $\chi_0$ for samples that deform at high temperatures, which is physically unacceptable. Based on the same observation as in \cite{LBL-10}, we assume that $\kappa _\rho$ is independent of strain rate and temperature, while $\kappa_\chi=\kappa_0\exp(T/c_1)$, where $\kappa_\rho $, $\kappa_0$ and $c_1$ are material constants. This is the last modification of the LBL-theory.

\section{Rescaled governing equations}
\label{Rescaled}

\begin{figure}[htb]
\centering \includegraphics[width=8cm]{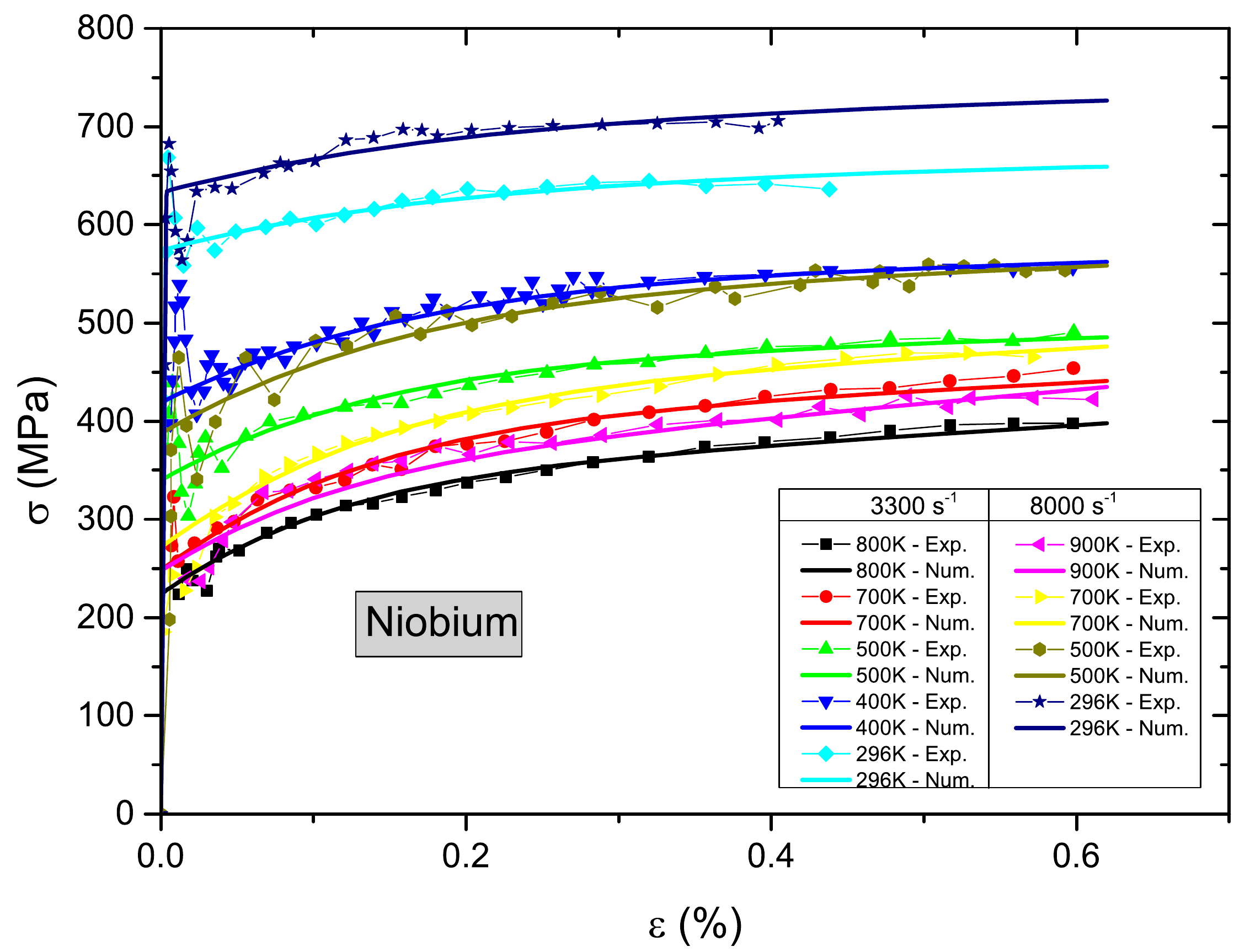} \caption{(Color online) Stress-strain curves for niobium at two different strain rates and six different temperatures: (i) $\dot\varepsilon = 3300\,$s$^{-1}$, $T=800\,$K (case 1); (ii) $\dot\varepsilon = 3300\,$s$^{-1}$,  $T=700\,$K (case 2); (iii) $\dot\varepsilon = 3300\,$s$^{-1}$,  $T=500\,$K (case 3), (iv) $\dot\varepsilon = 3300\,$s$^{-1}$, $T=400\,$K (case 4); (v) $\dot\varepsilon = 3300\,$s$^{-1}$,  $T=296\,$K (case 5); (vi) $\dot\varepsilon = 8000\,$s$^{-1}$,  $T=900\,$K (case 6), (vii) $\dot\varepsilon = 8000\,$s$^{-1}$, $T=700\,$K (case 7); (viii) $\dot\varepsilon = 8000\,$s$^{-1}$,  $T=500\,$K (case 8); (ix) $\dot\varepsilon = 8000\,$s$^{-1}$,  $T=296\,$K (case 9).  The experimental points are taken from Nemat-Nasser and Guo \cite{Nemat-Nasser2000}} \label{fig:niobium}
\end{figure}

\begin{figure}[htb]
\centering \includegraphics[width=8cm]{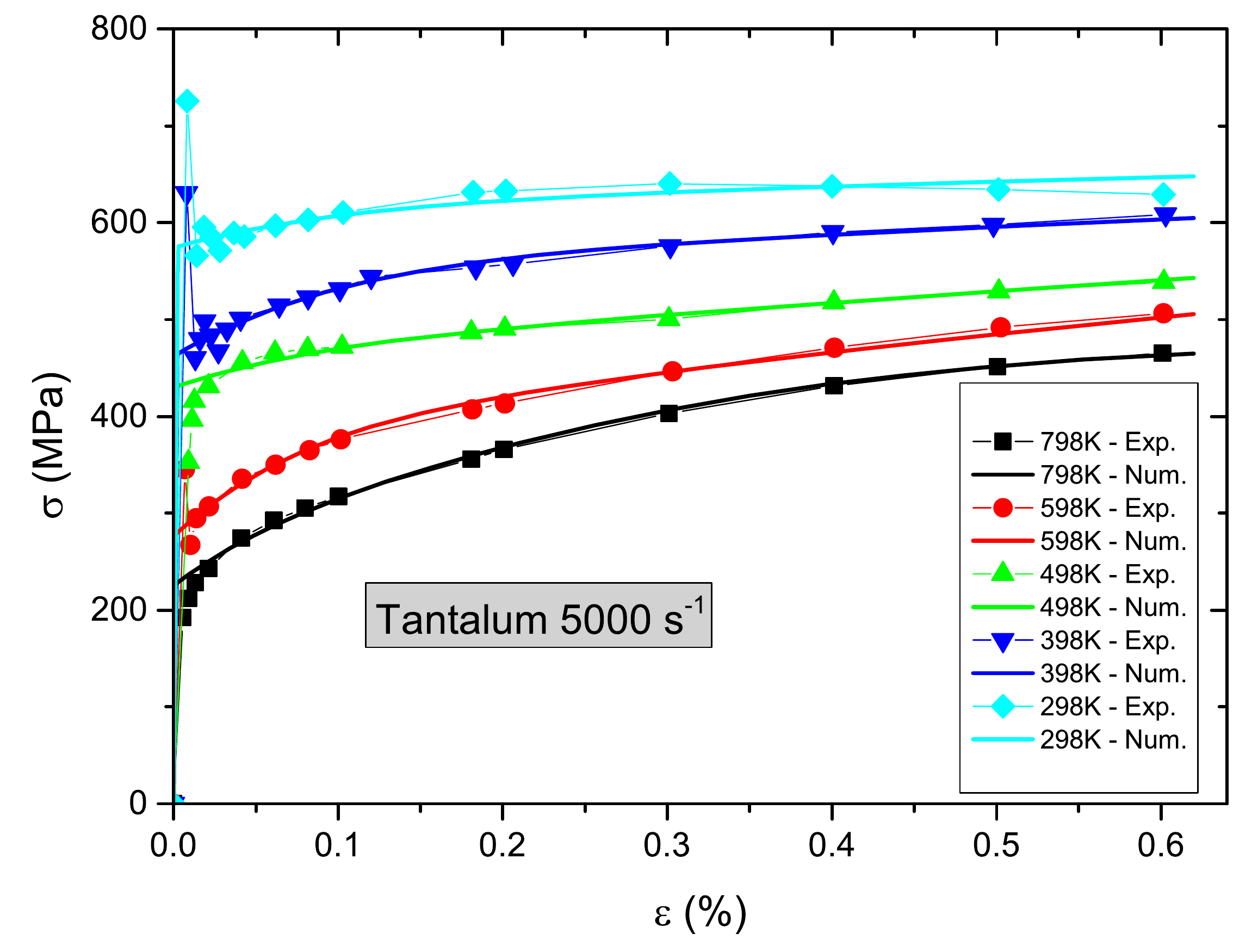} \caption{(Color online) Stress-strain curves for tantalum at the strain rate $5000/$s and five different temperatures: (i) $T=798\,$K (case 1); (ii) $T=598\,$K (case 2); (iii) $T=498\,$K (case 3), (iv) $T=398\,$K (case 4), (v) $T=298\,$K (case 5). The experimental points are taken from Nemat-Nasser and Isaacs \cite{Nemat-Nasser1997}} \label{fig:tantalum}
\end{figure}

For the purpose of numerical integration and parameter identification it is convenient to introduce the rescaled variables and rewrite the system \eqref{eq:stressrate} and \eqref{eq:15} in the dimensionless form. To this end, the dimensionless dislocation density and effective disorder temperature are introduced as
\begin{equation*}
\tilde{\rho }=a^2 \rho, \quad \tilde{\chi }=\frac{\chi }{e_d}.
\end{equation*}
With these rescaled quantities the dimensionless steady-state dislocation density at fixed configurational temperature becomes
\begin{equation*}
\tilde{\rho}_s(\tilde{\chi})=e^{-1/\tilde{\chi }}.
\end{equation*}
Let the dimensionless ordinary temperature be $\tilde{\theta} =T/T_P$. Then the normalized plastic strain rate can be written as
\begin{equation*}
q(\sigma,\rho,T)=\frac{2}{\pi}\dot{\varepsilon}^pt_0=(b/a)\tilde{q}(\sigma,\tilde{\rho},\tilde{\theta}),
\end{equation*}
where
\begin{equation*}
\tilde{q}(\sigma,\tilde{\rho},\tilde{\theta})=\sqrt{\tilde{\rho}}\exp \Bigl[-\frac{1}{\tilde{\theta}} e^{-\sigma/\tilde{\mu}_T\sqrt{\tilde{\rho}}} \Bigr] ,\quad \tilde{\mu}_T=(b/a)\bar{\mu}_T=\frac{3\pi}{2}(b/a)\mu_T.
\end{equation*}
The formula for $\zeta$ becomes
\begin{equation*}
\tilde{\zeta}(\tilde{\rho},\tilde{q}_0,\tilde{\theta})=\ln \Bigl( \frac{1}{\tilde{\theta}}\Bigr)-\ln\Bigl[\ln\Bigl(\frac{\sqrt{\tilde{\rho}}}{\tilde{q}_0}\Bigr)\Bigr] . 
\end{equation*}
We assume that $\tilde{\mu}_T$ scales like $\mu$ as a function of temperature $\tilde{\mu}_T(\tilde{\theta})=r\mu(\tilde{\theta})$, with $r$ being a material constant. Using $\tilde{q}$ instead of $q$ as the dimensionless measure of mean plastic slip rate, we are effectively rescaling $t_0$ by a factor $\pi b/2a$, $t_0=\tilde{t}_0\pi b/2a$, so that $\dot{\varepsilon}^p=\tilde{q}/\tilde{t}_0$. We also define $\tilde{q}_0=\dot{\varepsilon}\tilde{t}_0$. The parameter $\tilde{t}_0$ will not be assumed, but will be identified from the large scaled least squares analysis.

In terms of the introduced rescaled variables the governing equations read
\begin{align}
\dv{\sigma}{\varepsilon}&=2\mu (1+\nu) \Bigl( 1-\frac{\tilde{q}}{\tilde{q}_0}\Bigr), \notag
\\
\dv{\tilde{\rho}}{\varepsilon}&=K_\rho \frac{\sigma }{\tilde{\mu }_T \tilde{\zeta}^2(\tilde{\rho},\tilde{q}_0,\tilde{\theta})}\frac{\tilde{q}}{\tilde{q}_0}\Bigl[1-\frac{\tilde{\rho}}{\tilde{\rho}_s(\tilde{\chi})}\Bigr] , \label{eq:16}
\\
\dv{\tilde{\chi}}{\varepsilon}&=K_\chi \frac{\sigma }{\tilde{\mu}_T}\frac{\tilde{q}}{\tilde{q}_0} \Bigl(1-\frac{\tilde{\chi}}{\tilde{\chi}_0}\Bigr) ,\notag
\end{align}
where $K_\rho =\frac{\kappa_\rho a}{b}$ and $K_\chi = \frac{\kappa_\chi b}{a e_d}$. As discussed in the previous Section, we assume that $K_\rho$ is independent of the strain rate and temperature, while $K_\chi=c_0\exp(T/c_1)$, where $K_\rho $, $c_0$ and $c_1$ are material constants.

\section{Data analysis}
\label{DATA}

\begin{figure}[htb]
\centering \includegraphics[width=8cm]{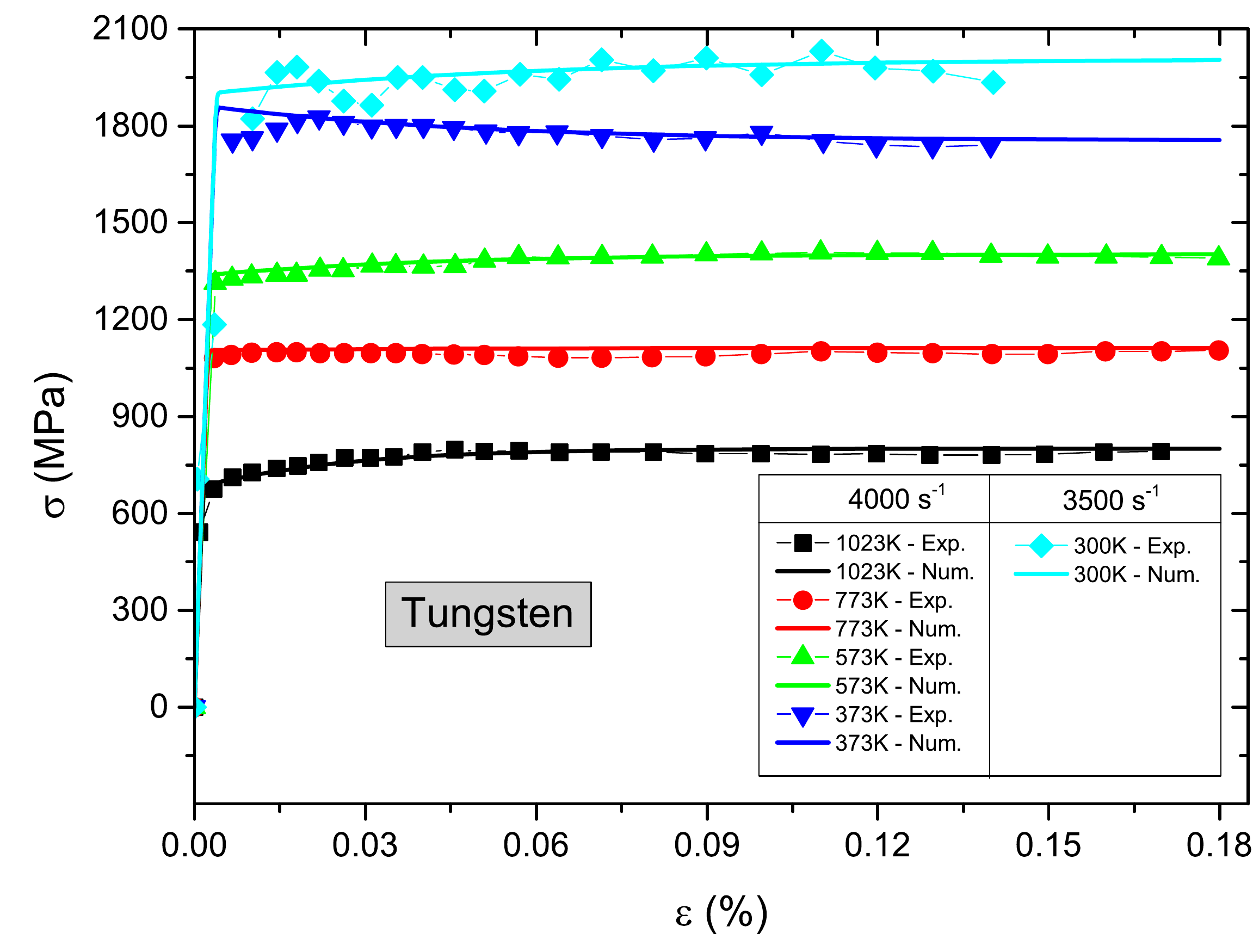} \caption{(Color online) Stress-strain curves for tungsten at two strain rates and five different temperatures: (i) $\dot\varepsilon = 4000\,$s$^{-1}$, $T=1023\,$K (case 1); (ii) $\dot\varepsilon = 4000\,$s$^{-1}$,  $T=773\,$K (case 2); (iii) $\dot\varepsilon = 4000\,$s$^{-1}$,  $T=573\,$K (case 3), (iv) $\dot\varepsilon = 4000\,$s$^{-1}$, $T=373\,$K (case 4); (v) $\dot\varepsilon = 3500\,$s$^{-1}$,  $T=300\,$K (case 5).  The experimental points are taken from Lennon and Ramesh \cite{Lennon2000}} \label{fig:tungsten}
\end{figure}

\begin{figure}[htb]
\centering \includegraphics[width=8cm]{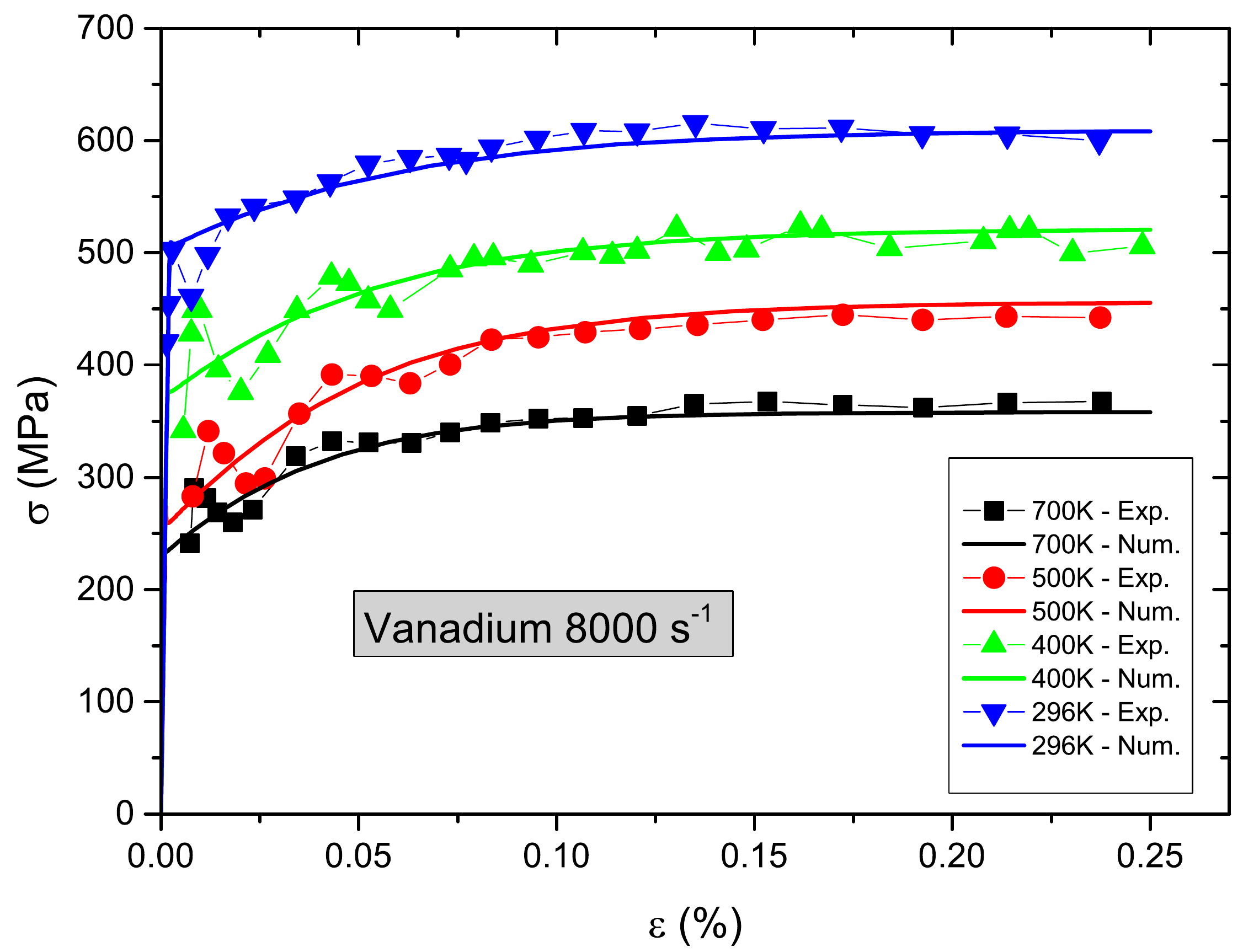} \caption{(Color online) Stress-strain curves for vanadium at the strain rate $8000/$s and four different temperatures: (i) $T=700\,$K (case 1); (ii) $T=500\,$K (case 2); (iii) $T=400\,$K (case 3), (iv) $T=296\,$K (case 4).  The experimental points are taken from Nemat-Nasser and Guo \cite{Nemat-Nasser2000a}} \label{fig:vanadium}
\end{figure}

The experimental results of Nemat-Nasser and Guo \cite{Nemat-Nasser2000} for niobium, Nemat-Nasser and Isaacs \cite{Nemat-Nasser1997} for tantalum, Lennon and Ramesh \cite{Lennon2000} for tungsten, and Nemat-Nasser and Guo \cite{Nemat-Nasser2000a} for vanadium, along with our theoretical results based on the equations of motion \eqref{eq:16}, are shown in Figs.~\ref{fig:niobium}, \ref{fig:tantalum}, \ref{fig:tungsten}, and \ref{fig:vanadium}, respectively. As discussed in \cite{LBL-10}, the kinetics of dislocation depinning breaks down at the very low temperature and small strain rate limits. This is the reason why the stress-strain curves obtained at quasi-static loading and low temperatures are not shown and analyzed. In order to compute the theoretical curves in these figures, we need values for seven system-specific parameters: the activation temperature $T_P$, the stress ratio $r$, the steady-state scaled effective temperature $\tilde\chi_0$, the dimensionless conversion factors $K_\rho$, $c_0$, $c_1$, and the characteristic time $\tilde{t}_0$. We also need initial values of the scaled dislocation density $\tilde\rho(\varepsilon = 0) \equiv \tilde\rho_i$, the effective temperature $\tilde\chi(\varepsilon = 0) \equiv \tilde\chi_i$, and the stress $\sigma_i$ which are determined by sample preparation. We take for granted that all specimens are initially stress-free, so $\sigma_i=0$ for all stress-strain curves. Further, we need a formula for the temperature dependent shear modulus $\mu(\tilde{\theta})$, which we take from \cite{Varshni70} to be 
\begin{equation*}
\mu(\tilde\theta) = \mu_1 - \frac{D}{\exp(T_1/T_P\,\tilde\theta)-1},
\end{equation*}
where $\mu_1 = 50.08\,$GPa, $D = 0.0207\,$GPa, $T_1 = 15\,$K for niobium \cite{Follansbee2014}, $\mu_1 = 65.25\,$GPa, $D = 0.38\,$GPa, $T_1 = 40\,$K for tantalum \cite{Chen1996}, $\mu_1 = 159.5\,$GPa, $D = 33.69\,$GPa, $T_1 = 1217\,$K for tungsten \cite{Dummer1998}, and $\mu_1 = 68.98\,$GPa, $D = 0.41\,$GPa, $T_1 = 45\,$K for vanadium \cite{Follansbee2014}.  Finally, Poisson's ratios of these materials are: $\nu=0.4$ (niobium), $\nu=0.34$ (tantalum), $\nu=0.28$ (tungsten), $\nu=0.37$ (vanadium).

Let these unknown parameters be the components of a vector denoted by $\vb{P}$ which belongs to the multi-dimensional space of parameters. This vector will be identified from $N_e$ experimentally measured stress-strain curves (of $N_e$ specimens of the same materials subjected to $N_e$ different thermal and loading conditions) as follows. Provided $\vb{P}$ is known, then we integrate the system \eqref{eq:16} at $N_e$ different temperatures and strain rates and $N_e$ different sets of initial conditions to find $N_e$ functions $\sigma _i(\varepsilon ,\vb{P})$, $i=1,\ldots ,N_e$, that depends on these parameters. Then we form the sum of squares
\begin{displaymath}
h(\vb{P})=\sum_{i=1}^{N_e}\sum_{j=1}^{N_i} (\sigma _i(\varepsilon _{(ij)},\vb{P})-\sigma _{(ij)})^2,
\end{displaymath}
where $(\varepsilon _{(ij)}, \sigma _{(ij)})$, $i=1,\ldots ,N_e$, $j=1,\ldots N_i$ correspond to the strains and stresses measured in experiment at $N_e$ different thermal and loading conditions (the index $j$ runs from 1 to $N_i$, with $N_i$ being the number of selected points on the curve $i$). We then find $\vb{P}$ by minimizing function $h(\vb{P})$ in the space of parameters subjected to physically reasonable constraints. Because the system of ordinary differential equations \eqref{eq:16} is stiff, we use the Matlab-solver ode15s for its numerical integration. Concerning the minimization of function $h(\vb{P})$: Since we want to find the global minimum (least squares), the best numerical package for this is the Matlab-GlobalSearch minimization.

	\begin{table}
	\begin{center}
		\begin{tabular}{ | l | l | l |l | l |}
			\hline
			\diagbox[width=10em]{Parameter}{Material} & niobium    & tantalum   & tungsten   & vanadium   \\ \hline
			$T_{P}$      & 38380      & 32360      & 35890      & 45000      \\ \hline
			$r$          & 0.146    & 0.127    & 0.133    & 0.024   \\ \hline
			$\tilde{\chi}_0 $      & 0.228    & 0.199    & 0.18    & 0.282    \\ \hline
			$c_{0}$      & 0.0167   & 0.0176   & 0.0011  & 148     \\ \hline
			$c_{1}$      & 187     & 114     & 208     & 169     \\ \hline
			$K_{\rho}$  & 0.527    & 0.801     & 1.789     & 7.367     \\ \hline
			$\tilde{t}_{0}$        & 3.68e-13 & 2.e-13 & 3.78e-13 & 3.6e-12 \\ \hline
		\end{tabular}
		\caption{\label{tab:TDTParameters}Physics-based material parameters for bcc-crystals.}
	\end{center}
	\end{table}

The finding of initial guesses and reasonable constraints for the parameters is not easy. In some cases  deliberations of physical nature can help. For instance, the estimation of $\tilde{\chi}_0$, given in \cite{LBL-10}, is based on the following argument. The definition of $\tilde{\chi}_0$ can be interpreted very roughly as a system-independent geometric criterion, weakly analogous to the idea that amorphous materials become glassy when their densities are of the order of maximally random jammed packings, or to the Lindemann criterion \cite{Lindemann1910} according to which crystals melt when thermal vibration amplitudes are of the order of a tenth of the lattice spacing. In that spirit, we guess that $\tilde{\chi}_0$ is the dimensionless effective temperature at which the spacing between dislocations is roughly $10a$, or about 100 atomic spacings. Thus we guess that $1/\tilde{\chi}_0\sim 2\ln 10\sim 4$ and, therefore $\tilde{\chi}_0\sim 0.25$. Then, with two experimentally measured (or estimated) steady-state stresses we can find the rough estimates of $T_P$, $r$ (see the details in \cite{Le2020}). Similar estimations for the conversion factors and the initial conditions can be found in \cite{Le2020}. For the microscopic time $\tilde{t}_0$ we use the initial estimation $\tilde{t}_0\sim 10^{-12}$s.

	\begin{table}
	\begin{tabular}{|l|c|c|c|c|}\hline
		\diagbox[width=10em]{Case}{Material} & niobium &	tantalum	& tungsten & vanadium \\ \hline
		\multirow{2}{*}{case 1}		& 0.166	&	0.167	&	0.177	&	0.138 \\ 
									& 0.00093	&	0.00133	&	0.00254	&	0.01204 \\ \hline
		\multirow{2}{*}{case 2}		& 0.167	&	0.164	&	0.174	&	0.107 \\
									& 0.00089	&	0.00103	&	0.00317	&	0.01121 \\ \hline
		\multirow{2}{*}{case 3}		& 0.16	&	0.164	&	0.17	&	0.194 \\
									& 0.00099	&	0.00182	&	0.00256	&	0.01481 \\ \hline
		\multirow{2}{*}{case 4}		& 0.161	&	0.164	&	0.165	&	0.206 \\
			      					& 0.00114	&	0.00148	&	0.00265	&	0.02008 \\ \hline
		\multirow{2}{*}{case 5}		& 0.161	&	0.159	&	0.166	&		\\ 
									& 0.00155	&	0.00157	&	0.00218	&		\\ \hline
		\multirow{2}{*}{case 6}		& 0.173	&		&		&		\\ 
									& 0.00132	&		&		&		\\ \hline
		\multirow{2}{*}{case 7}		& 0.169	&		&		&		\\ 
									& 0.00098	&		&		&		\\ \hline
		\multirow{2}{*}{case 8}		& 0.166	&		&		&		\\ 
									& 0.00121	&		&		&		\\ \hline
		\multirow{2}{*}{case 9}		& 0.165	&		&		&		\\ 
									& 0.00180	&		&		&		\\ \hline
	\end{tabular}
	\caption{\label{tab:TDTInital} The initial dimensionless configurational temperature $\tilde{\chi}_i$ and dislocation density $\tilde{\rho}_i$ of $N_e$ specimens subjected to $N_e$ compression tests ($N_e$ varies from four to nine depending on the material). In each cell the upper number corresponds to $\tilde{\chi}_i$, while the lower number to $\tilde{\rho}_i$. For the cases see the Figures and their captions.}
	\end{table}	

With these guesses and constraints the large scale least squares method \cite{LTL-17,LT-17} enables one to identify the seven basic parameters and the initial values of $\tilde{\rho}$ and $\tilde{\chi}$. The identified parameters turn out robust and are presented in Table~\ref{tab:TDTParameters} for the above materials. We also found the initial conditions for the samples of these materials subjected to compression testing. Table~\ref{tab:TDTInital} represent these identified initial data, where in each cell the upper number corresponds to $\tilde{\chi}_i$, while the lower number to $\tilde{\rho}_i$. The cases presented in this table are explained in the legends and captions of Figs.~\ref{fig:niobium}-\ref{fig:vanadium}.

The agreement between theory and experiment seems to be within the range of experimental uncertainties. There are only a few visible discrepancies. The first of these is the absence of the sharp yield points that are clearly seen in case of niobium, tantalum, and vanadium. Although the upper yield points followed by a stress drops can be simulated by choosing a low initial $\chi_i$, we have not been able to find the optimal set of parameters that provides the sharp yield points and the correct behavior thereafter for all stress-strain curves. One of the possible reason could be the material instability that affects the numerical analysis. Perhaps, the particular dislocation removal mechanism in bcc crystals should be taken into account to explain the presence of the sharp yield points. Another observable discrepancy concerns the oscillation of the stress-strain curves, particularly at low temperatures, which is not present in the theoretical curves. For example, the experimental data in Figure~\ref{fig:tungsten} for $\dot\varepsilon =3500/$s and $T=300\,$K show a small oscillation in the stress-strain curve. Similarly, when deformed at high strain rates, the theoretical stress-strain curves for niobium and vanadium shown in Figure~\ref{fig:niobium} and Figure~\ref{fig:vanadium} deviate  from the real experimental oscillating curve. To clarify this deviation as well as a small softening behavior at large strains, which are possibly due to the thermal effect, it would be useful to extend the TDT and to include the equation of motion for the kinetic-vibrational temperature \cite{LTL-17}. The last discrepancy concerns the absence of the strain bursts which are due to the acoustic emission during the motion of dislocations \cite{Csikor2007}. It is also interesting to mention an unusually large value of the conversion factor $c_0$ for vanadium, which may indicate the rapidly changing complexity of the dislocation network, causing the configuration entropy to rapidly approach its stationary value $\tilde{\chi}_0$.

\section{Conclusion}
\label{Conclusions}
On the whole, these results seem to us to be quite satisfactory. We did not know at the beginning of this study whether the thermodynamic dislocation theory is applicable to bcc-crystals which may have quite different mechanisms of dislocation depinning as well as the strong influence of the cross slip. Our results of simulations let us conclude that TDT can indeed be applied to this class of materials. Furthermore, we have found the physics-based parameters that enables one to predict the stress-strain curves in a wide range of temperatures and strain rates. One interesting problem remains still however open: How to predict the strong load drops at the onset of adiabatic shear banding as shown on Figure 4 in \cite{Nemat-Nasser2000a}. To solve this problem, the theory must be extended to include the interaction between the configurational and kinetic-vibrational subsystems in the spirit of \cite{LTL-18}.

\end{document}